# Specific features of the average magnitudes and luminosities of quasars and galaxies as a function of redshift and their interpretation in the modified cosmological model

Iurii Kudriavtcev

*In this article we examine the dependency of the magnitude of quasars from SDSS DR-5 as a function of redshift in different frequency ranges in the system (u,g,r,i,z). We show that on the smoothed curves mag(Z) in frequency ranges u,g,r there are characteristic features both similar for all the curves and different for different curves. At Z<2 all the curves have practically the same form with two characteristic sections of the negative slope, at Z>2 the forms of the curves significantly differ. The nature of these differences can be interpreted as an influence of light absorption by the neutral hydrogen on Lyman-alpha line, manifesting on the curves in the ranges u,g and missing on the curve in the range r.*

*Then we have studied the dependencies on redshift of the averaged over the sky magnitude values in the range r, free from the Lyman-alpha absorption effect, of the galaxies from catalogs 2DFGRS, Millenium, Gyulzadian and quasars from SDSS DR-7. We have shown that for all the objects, both quasars and galaxies, the dependencies of the magnitudes average values on Z are close by values and form and have a nature of converging oscillations with areas of negative slope near Z ≈ 1 and Z ≈ 2, corresponding to the increase of the apparent luminosity with the increasing of the distance.*

*We have calculated the corresponding values of the average absolute luminosities for the flat space and different variants of the ΛCDM model. For all the used calculation variants the average luminosities of the objects increase with the growth of the redshift. When the redshift increases from 0.02 to 0.2, the average radiation power of the galaxy increases by 40-50 times, and when the first one increases from 0.02 to 2, then the second one increases by thousands times. Such changes appear to be impossible which leads us to the assumption that these results are related to the used calculation methods and the features of the metric tensor underlying of the standard cosmological model.*

*In the modified cosmological model which built on the metrics obtained with the considering the non-zero value of the differential of scale factor, appear features that are missing in the standard model, including the effect of increase of the apparent luminosity with the increase of distance. We study the possibility of the numerical modeling of the dependency of magnitudes of the distant radiating objects on redshift within the modified model, using the methodology built on the beliefs about the ray's movement from the source to the observer in a space with randomly changing curvature. We have shown that the calculation performed with this methodology allows obtaining a curves close to dependencies of average magnitudes on redshift issuing from observation data.*

**98.80.-k**

## 1. Introduction.

This paper continues to analyze data on redshift and luminosity of quasars from SDSS-DR5 catalog [1], which were previously used to study the opposite pairs of quasars in order to confirm the phenomenon of central symmetry of the celestial sphere [2]. Then the results obtained are verified with expanding to a more wide groups of distant observed objects - quasars from SDSS-DR7 catalog [3] and galaxies from the 2dF Galaxy Redshift Survey [4], the Millennium Galaxy Catalogue [5] and the Second Byurakan Survey [6].

## 2. Average magnitudes of quasars as a function of redshift

We analyzed here the quasars of the subequatorial area of the celestial sphere ($-1.4^0<DE<1.4^0$) - about 12900 quasars from the catalog SDSS DR-5. The Sloan Digital Sky Survey (SDSS) [1], contains the data about the visible luminosity of the quasars expressed in the modified system of stellar magnitudes ASINH [7], related to the values of the visible luminosity l(Z) by the expression:

$$\text{mag} = -2.5/\ln(31)*[\text{asinh}((f/f_0)/(2b)) + \ln(b)]. \qquad (1)$$

where $f \equiv l(Z)$, $f_0$ — classical zero point of the magnitude scale, b - adjusting coefficient, given for the different frequency ranges (u,g,r,i,z) in the system [7].

On the Picture 1 we see diagrams mag(Z) in the ranges u[300...400nm], g[400...500nm] and r[600...750nm], according to the data from the SDSS-DR5 quasar catalog.

Picture 1A shows the diagram umag(Z) without smoothing. In relation to the big variance of the individual characteristics it is harder to see the regularities of the formation of the dependency, on the Pictures 1B, 1C, 1D we show the dependencies smoothed by means of the averaging of the magnitudes by 100 following quasars placed in the order of ascending of Z:

$$\text{mag}_i\_\text{mean}(100) = (\text{mag}_i + \text{mag}_{i+1} + ... + \text{mag}_{i+100})/100. \qquad (2)$$

where $1<i<12800$. There is a practically horizontal area with mild pulsations with the maxima at $Z \approx 0,5$ and $Z \approx 1,5$ and minima at $Z \approx 1$ and $Z \approx 2$ on all the dependencies shown on the Picture 1.

### 3. Effect of Lyman-alpha absorption

The curves umag(Z) and gmag(Z) at big values of Z deviate from the flat and mildly periodical dependency and drastically increase. For umag the area of the drastic increasing begins at $Z \approx 2,5$, for gmag at $Z \approx 3$, on the curve rmag the increase area is absent. These particularities allow to suppose that this area of the drastic increase mag, corresponding to the visible luminosity, can be related to the phenomenon of the resonant absorption of the light in the interstellar space [8] on the frequencies, corresponding to the different values of Z. If there is a resonant absorption of the light on the definite wave length it is going to be noted by the observer at different values of Z for different ranges of observation. To check this assumption let us calculate the lengths of the light waves corresponding to the middles of the ranges «u» and «g», in the corresponding points of the increase areas of the curves umag(Z) and gmag(Z). In case if the increase of the curves is caused by the resonant absorption on the definite wave length equal for the both curves, the obtained wave lengths, defined by the length of the absorption wave, must coincide.

The length of the wave of resonant absorption $\lambda_{abs}$, visible by the observer at $Z_{abs}$, can be calculated by the formula

$$\lambda_{abs} \approx \lambda_c /(1+ Z_{abs}). \qquad (3)$$

Calculating for the high limits of ranges «u» and «g» and the same values mag = 21, we obtain:

$$\lambda_{absU} \approx 400/(1+ 2.7) \approx 108 \text{ nm.}$$

$$\lambda_{absG} \approx 500/(1+ 3.6) \approx 108 \text{ nm.}$$

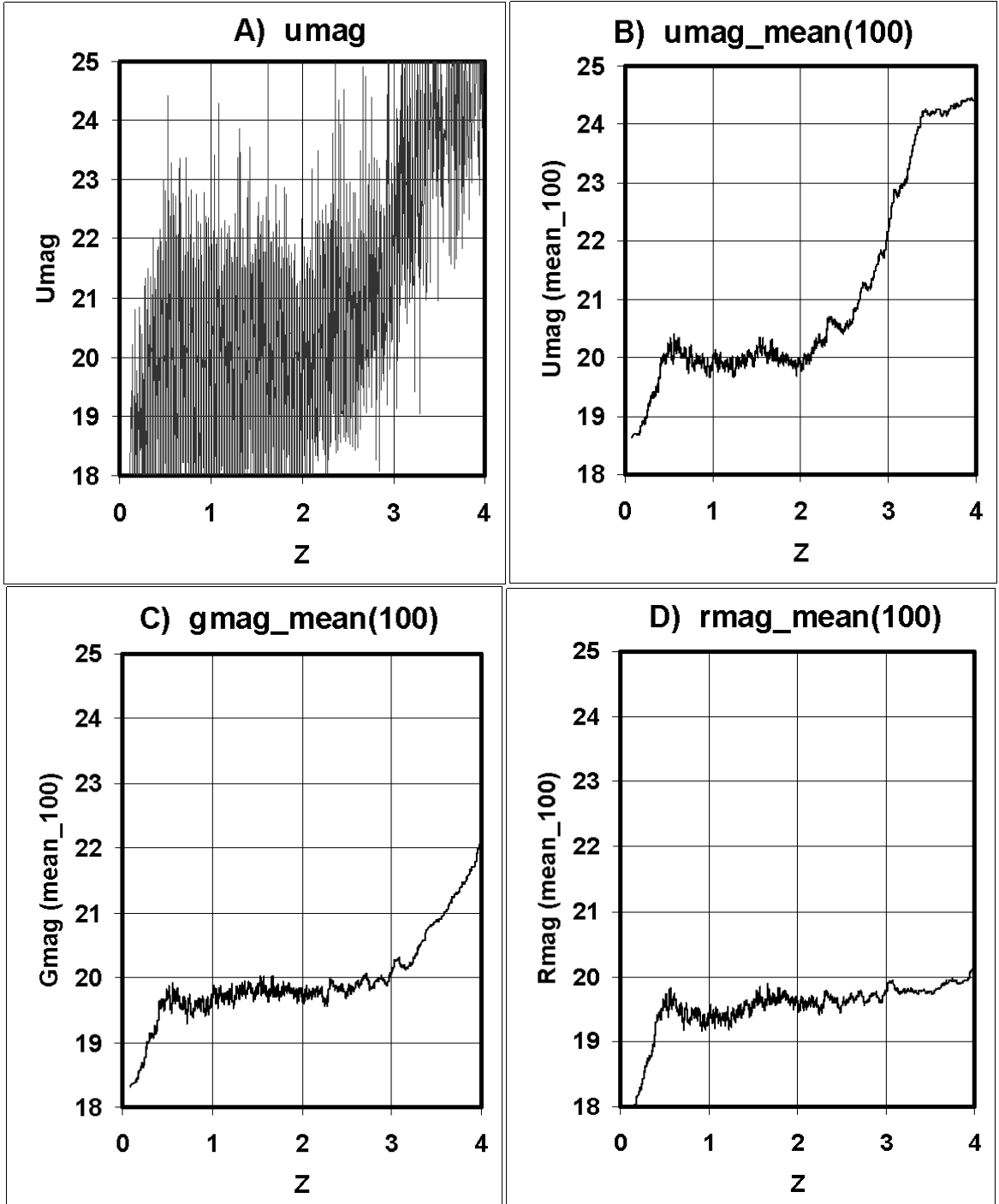

**Picture 1.** Dependencies of the stellar magnitudes umag(Z), gmag(Z), rmag(Z) of the quasars of the of the subequatorial area of the catalog SDSS (12800 quasars). A) — without smoothing, B),C),D) — smoothed by the averaging by 100 units [mean(100)].

This way $\lambda_{abs}u \approx \lambda_{abs}g$, which corresponds to the assumption about the formation of the increase area of the curves at the expense of the resonant absorption. Let us note that the obtained value of $\lambda_{abs}$ appears to be close to Lyman-alpha (121,5 nm.). If we take for evaluation the very beginning of the increase intervals of the curves (for instance, by making its linear extrapolation to the level of the horizontal area mag ≈ 20), we will obtain the value $\lambda_{abs,}$ that practically coincide with the absorption wave length of Lyman-alpha

$$\lambda_{absU} \approx 400/(1+2.3) \approx 121 \text{ nm.}$$

$\lambda_{absG} \approx 500/(1+3.1) \approx 122$ nm.

At that for the range «r» we will obtain the beginning of the increase area at $Z \approx 750/122\text{-}1\approx5.1$, which corresponds to its absence on the curve rmag(Z), Picture 1D.

### 4. Average luminosity of quasars as a function of their redshift

In the range «r» the effect of Lyman-alpha absorption is absent, which allow us to consider, that it reflects some common features of quasars.

The empiric curve, practically coinciding with the smoothed curve rmag(Z) shown on the Picture 1D, can be described by the expression:

$$l_r(Z) \sim L \exp(-k\chi)(1+q\chi)/(\sin^2\chi + q\chi); \qquad (4)$$

where L - absolute luminosity (power), $\chi \approx \pi Z$, k, q – constants.

On the Picture 2 we show the corresponding to the expressions (4) and (1) calculated curve for the values of the parameters, chosen on basis of its most exact correlation to the smoothed curve rmag(Z) (q=0.9, k=0.03).

Visible luminosity of the distant source in Euclidean space is related [8] to the some absolute luminosity L by the expression:

$$l = (L/4\pi r^2)(1+Z)^{-2}; \qquad (5)$$

where r — distance to the source, Z — its redshift.

Relative speed of the object that moves away is related to its redshift by the expression

$$Z = ((1+\beta)/(1-\beta))^{1/2} -1; \qquad (6)$$

from where for Z < 1, at which relativistic effects manifest rather mildly, we obtain

$$\beta \approx Z/(1+Z); \qquad (7)$$

(at Z<1 the measure of inaccuracy of this formula is less than 1/6 = 16,6 %, which is rather enough for the evaluation and obtaining the quality dependencies). Expressing velocity of moving away via the Hubble constant (v = $\beta$c = Hr ), we obtain

$$r = v/H = (c/H)(Z/1+Z); \qquad (8)$$

Inserting to (5), we obtain [8] the expression:

$$l(Z) = (H/c)^2(L/4\pi Z^2). \qquad (9)$$

Dependency mag(Z), corresponding to the expression (9), is also shown on the Picture 2.

The comparison of (4) and (9) shows that averaged values of the luminosity (power) of quasars are increasing with redshift, which correlates to the existing in the literature data about the extremely high luminosity of the quasars with big redshifts.

The given above empiric formula (4) allows to express this dependency numerically. Comparing right and left parts of the expressions (4) and (9), we obtain:

$$L(Z) = K\chi^2(1+q\chi)\exp(-k\chi)/(\sin^2\chi + q\chi); \qquad (10)$$

where K - constant, $\chi \approx \pi Z$, q=0.9, k=0.03.

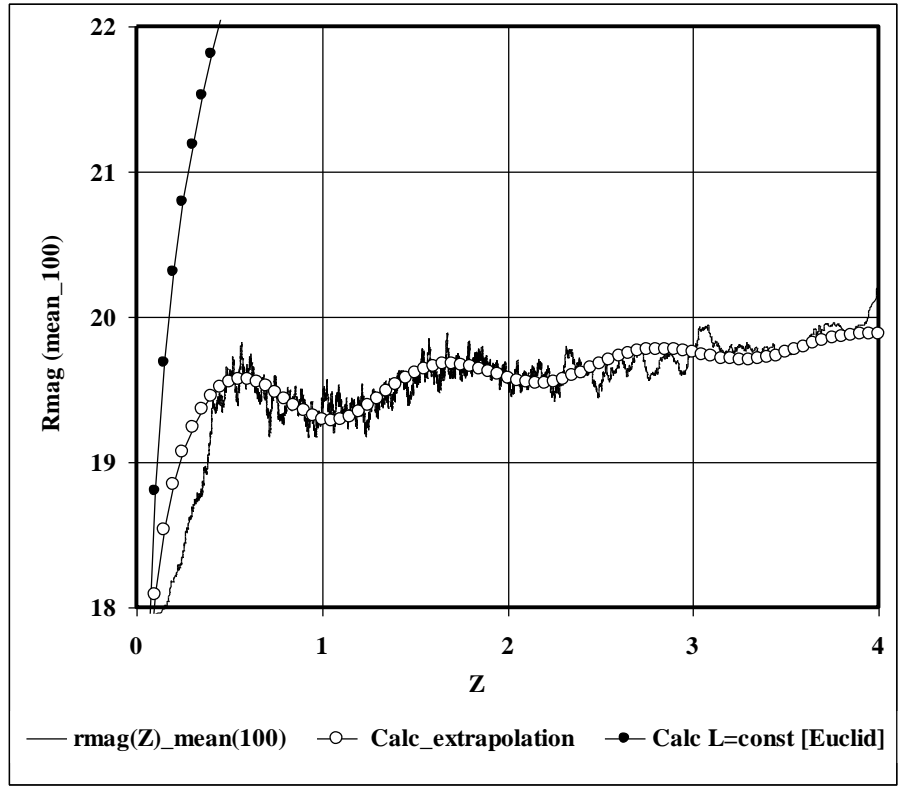

**Picture 2.** Comparison of the rmag(Z)_mean(100) with the empiric curve (4) for q=0.9, k=0.03 and with the dependency l(Z) for the source with the defined luminosity L in Euclidean space.

As formula (4) describes satisfactorily the dependency rmag(Z) only at $Z \geq 0.5$, and at $Z \rightarrow 0$ corresponds to the expression (9), relating the visible and absolute luminosity of the observed object only at q=0, constant K can be defined from the condition $Z \rightarrow 0$, q=0. From where K=L(0). Dependency calculated by the expression (10), is shown on the Picture 3.

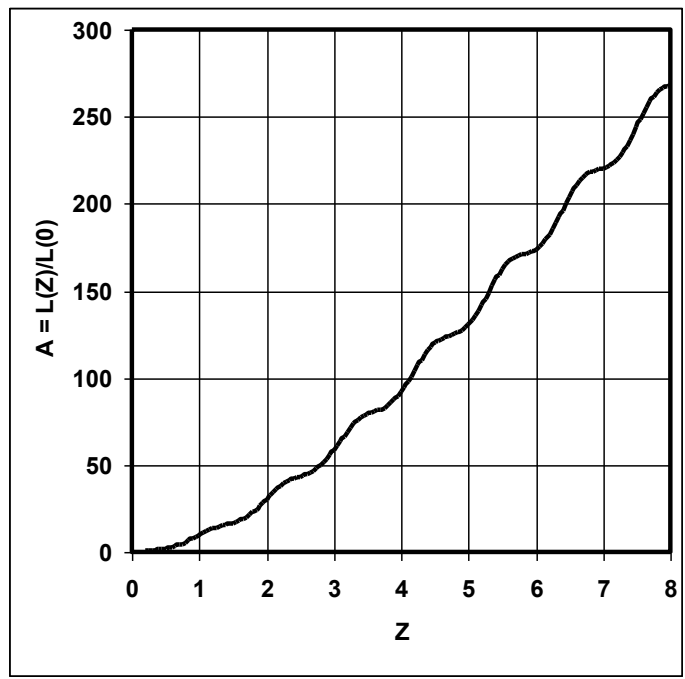

**Picture 3.** The curve A(Z) = L(Z)/L(0) according to (10).

Recently discovered quasar with Z=7.085 [9] has absolute luminosity $6.3 * 10^{13}L_{sun}$. From the curve on the Picture 3 we see that for $Z \approx 7$ relation $L(Z)/L(0) \approx 220$, which corresponds to the value $L(0) \approx 3*10^{11}L_{sun}$, close to the luminosity Milky Way ($\approx 10^{11}L_{sun}$).

Because of the fact that the negative slope of the magnitude dependency on redshift corresponds not to the decrease but increase of the apparent luminosity of the objects with the increase of the distance to them and because of the resulting dependence L(Z) we find it interesting to analyze this phenomenon more closely and expand it to a more wide group of other distant observed objects – the galaxies.

### 5. Averaged magnitudes of the galaxies from the catalogs «2DF GRS», «Millenium», «Gyulzadian» and quasars from SDSS DR-7.

Because of the big amount of the analyzed objects and very severe unevenness of the galaxies distribution by the redshift, the averaging of the magnitude was performed not by the given number of objects located according to the increasing of Z, as in [1], but by all the objects of the catalog in every interval of Z.

For the quasars from the catalog SDSS DR-7 (circa 105 thousands of objects) we have chosen a constant width of the interval $\Delta Z = 0.1$. The obtained dependency of the average values of rmag(Z) is shown on the Picture 4. The amount of quasars for interval (point of the curve) is from 300-500 to 5000-6000 units.

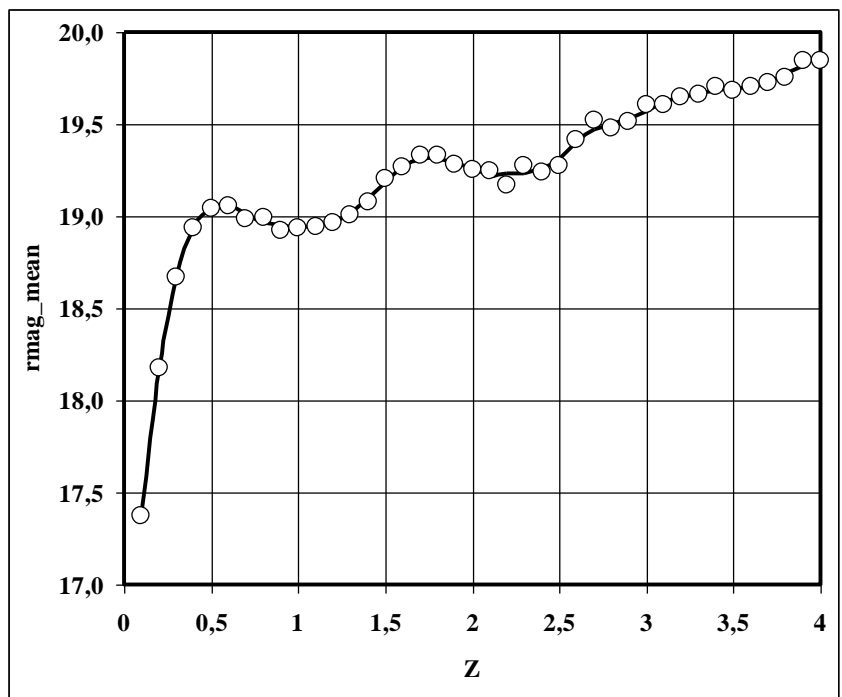

**Picture 4.** Dependency rmag_mean(Z) for 105 thousands of quasars from SDSS DR-7 [3].

To analyze the distributions by Z of the average galaxies magnitudes we used the data from the catalogs 2dF Galaxy Redshift Survey (2dFGRS Team, 1998-2003) [4] and Millennium Galaxy

Catalogue (Liske+, 2003) [5]. For the small values of Z we also used the data from the catalog Second Byurakan Survey galaxies. Optical database (Gyulzadian+, 2011) [6]. Because of the very uneven distribution of the galaxies by Z, the averaging of the galaxies magnitudes was done by the intervals of the width variable from $\Delta Z = \pm 0.005$ for $Z \approx 0.02$ with gradual increasing till $\Delta Z = \pm 0.3$ for $Z \approx 3$.

The obtained dependencies on Z of the average values of the magnitudes in r-range are shown on the Picture 5. The averaging parameters are shown in the Table 1.

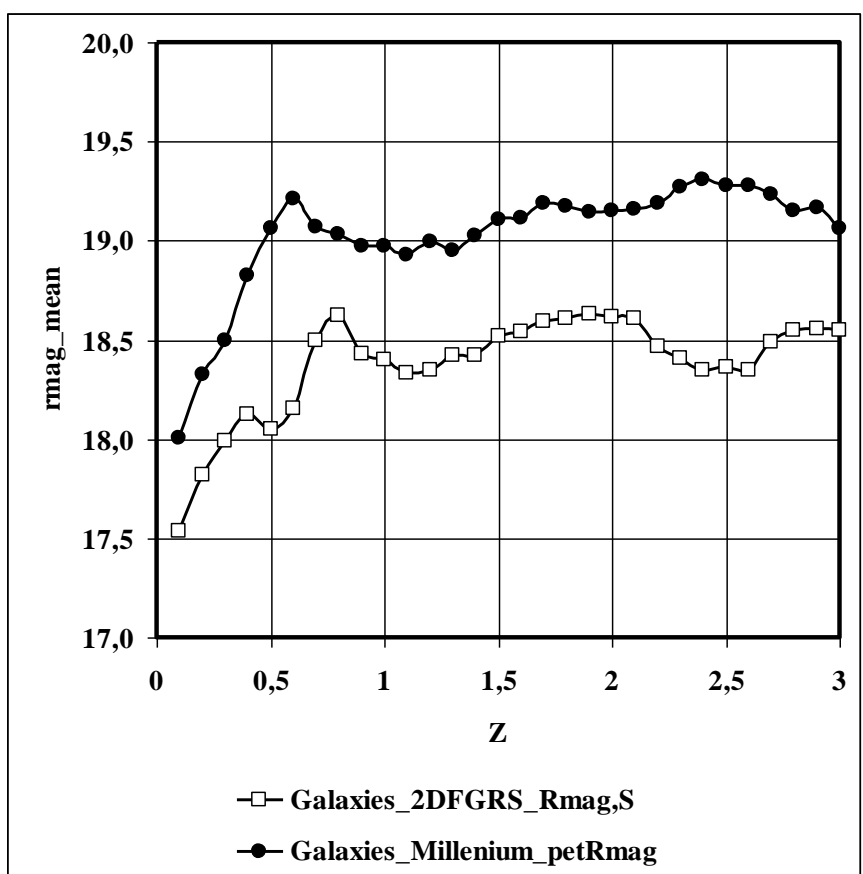

**Picture5.** Dependencies Rmag_mean(Z) for the galaxies from catalogs 2DF GRS [4], Millenium [5].

| | | **Galaxies 2DFGRS** | | **GalaxiesMillenium** | |
|---|---|---|---|---|---|
| **Z** | **Averaging interval Z1<Z<Z2** | **Number of galaxies in the interval** | **Average valueRmag,S** | **Number of galaxies in the interval** | **Average valuepetRmag** |
| 0,1 | 0,1—0,12 | 35367 | 17,54 | 1384 | 18,01 |
| 0,2 | 0,2—0,22 | 10129 | 17,82 | 485 | 18,32 |
| 0,3 | 0,28—0,32 | 1039 | 17,99 | 112 | 18,50 |
| 0,4 | 0,35—0,45 | 71 | 18,13 | 47 | 18,83 |
| 0,5 | 0,45—0,55 | 21 | 18,05 | 26 | 19,07 |
| 0,6 | 0,55—0,65 | 26 | 18,15 | 28 | 19,21 |
| 0,7 | 0,6—0,8 | 12 | 18,50 | 49 | 19,07 |
| 0,8 | 0,7—0,9 | 8 | 18,63 | 56 | 19,03 |
| 0,9 | 0,7—1,1 | 21 | 18,43 | 116 | 18,98 |
| 1 | 0,8—1,2 | 20 | 18,40 | 126 | 18,97 |
| 1,1 | 0,9—1,3 | 18 | 18,33 | 129 | 18,93 |
| 1,2 | 1,0—1,4 | 20 | 18,35 | 127 | 19,00 |
| 1,3 | 1,1—1,5 | 13 | 18,42 | 129 | 18,95 |
| 1,4 | 1,2—1,6 | 14 | 18,42 | 141 | 19,03 |

| 1,5 | 1,3—1,7 | 15 | 18,52 | 133 | 19,10 |
|---|---|---|---|---|---|
| 1,6 | 1,4—1,8 | 18 | 18,54 | 137 | 19,11 |
| 1,7 | 1,5—1,9 | 17 | 18,59 | 129 | 19,19 |
| 1,8 | 1,6—2 | 15 | 18,61 | 116 | 19,18 |
| 1,9 | 1,7—2,1 | 17 | 18,63 | 122 | 19,14 |
| 2 | 1,8—2,2 | 16 | 18,61 | 113 | 19,15 |
| 2,1 | 1,9—2,3 | 17 | 18,61 | 97 | 19,16 |
| 2,2 | 2—2,4 | 20 | 18,47 | 81 | 19,19 |
| 2,3 | 2,1—2,5 | 20 | 18,41 | 62 | 19,27 |
| 2,4 | 2,2—2,6 | 17 | 18,35 | 53 | 19,31 |
| 2,5 | 2,3—2,7 | 13 | 18,36 | 51 | 19,28 |
| 2,6 | 2,4—2,8 | 12 | 18,35 | 42 | 19,28 |
| 2,7 | 2,4—3 | 15 | 18,49 | 48 | 19,23 |
| 2,8 | 2,5—3,1 | 12 | 18,55 | 33 | 19,15 |
| 2,9 | 2,6—3,2 | 11 | 18,55 | 21 | 19,17 |
| 3 | 2,7—3,3 | 8 | 18,55 | 13 | 19,06 |

**Table 1.** Obtaining parameters of the magnitudes average values of the catalogs 2DF GRS [4] and Millenium [5], shown on the Picture 5.

Studied more closely, dependencies for small Z are shown on the Picture 6 with the averaging parameters brought in the Table 3.

The use of the "average magnitude value" notion is legit which is proved by the shown on the Picture 7 distributions of the galaxies from the catalog  2DF GRS [4] by values of the magnitudes Rmag,Sfor the ranges Z, corresponding to the marked cells of the Table 2. The distributions have a smooth bell-shaped form, close to the normal distribution. Locations of the average values of the magnitudes from the Table 2 are marked with crosses.

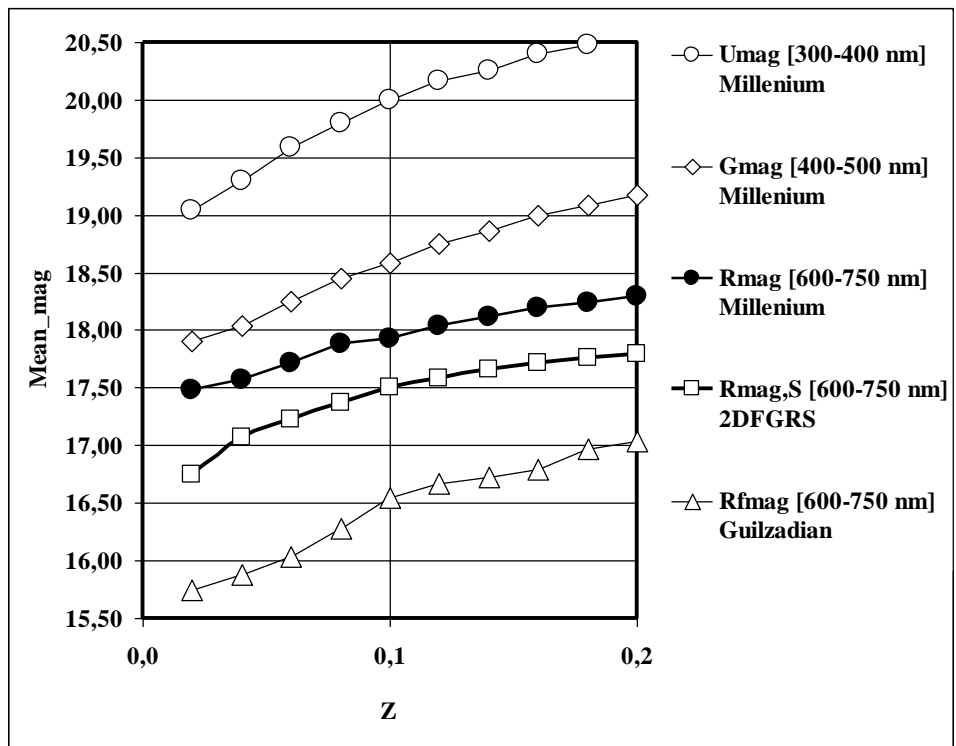

**Picture 6.** Dependencies Rmag_mean(Z) for the galaxies from the catalogs 2DF GRS [4], Millenium [5] and Gyulzadian [6] for the small values of Z.

| Z | Millenium | | | | 2DFGRS | | Guilzadian | |
|---|---|---|---|---|---|---|---|---|
| | N | Umag | Gmag | Rmag | N | Rmag,S | N | Rfmag |
| 0 | | | | | | | | |
| 0,02 | 269 | 19,04 | 17,91 | 17,48 | 3890 | 16,75 | 419 | 15,74 |
| 0,04 | 569 | 19,30 | 18,04 | 17,57 | 6096 | 17,07 | 480 | 15,88 |
| 0,06 | 698 | 19,59 | 18,25 | 17,72 | 14267 | 17,22 | 256 | 16,04 |
| 0,08 | 1530 | 19,80 | 18,46 | 17,88 | **17266** | 17,37 | 149 | 16,28 |
| 0,10 | 1773 | 19,99 | 18,58 | 17,92 | 14384 | 17,50 | 61 | 16,54 |
| 0,12 | 1431 | 20,16 | 18,75 | 18,04 | 15289 | 17,59 | 32 | 16,67 |
| 0,14 | 1457 | 20,26 | 18,87 | 18,12 | 13471 | 17,66 | 22 | 16,72 |
| 0,16 | 905 | 20,40 | 19,00 | 18,20 | 9860 | 17,72 | 22 | 16,79 |
| 0,18 | 991 | 20,48 | 19,08 | 18,24 | **7378** | 17,76 | 8 | 16,98 |
| 0,20 | 473 | 20,56 | 19,18 | 18,29 | 5755 | 17,79 | 8 | 17,04 |

**Table 2.** Obtaining parameters of the magnitudes average values of the quasars from the catalogs 2DF GRS [4], Millenium [5], and Guilzadian [6], for small values of Z, shown on the Picture 6. Averaging intervals $\Delta Z = \pm\ 0.005$.

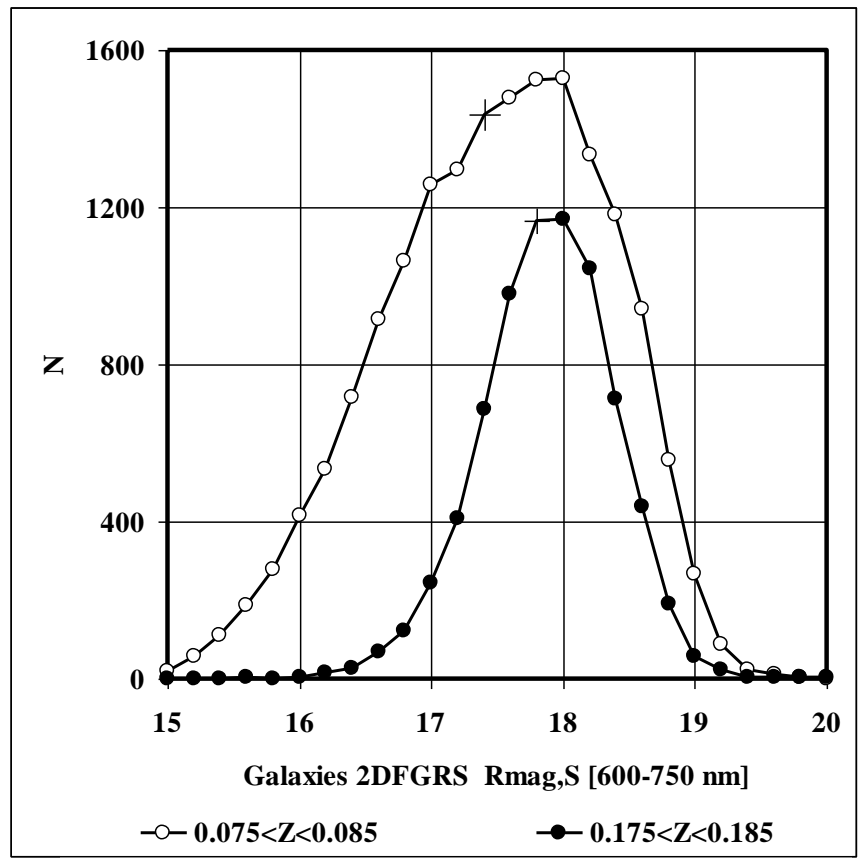

**Picture 7.** Distribution of the galaxies from the catalog2DF GRS [4] by values Rmag,S for the galaxies in the intervals Z, corresponding to the marked cells of the Table 3.

The data on the Pictures 5,6, shows that the dependencies on Z of the average magnitudes values of all the distant radiating objects, both quasars and galaxies, have the converging oscillations nature with areas of negative slope $Z \approx 1$ and $Z \approx 2$, corresponding to the increase of the apparent luminosity of the object with increase of the distance.

### 6. Calculation of absolute luminosities corresponding to the average magnitudes.

In Friedman's standard cosmological model the light from the distant object is distributed in all the directions in a flat and almost flat space. For the flat Euclidean space the apparent luminosity of the radiating object l is related to the absolute luminosity L (radiated power) [8] by the expression (9)and distance d to the object is equal

$$d = (c/H_0)Z; \qquad (11)$$

where $H_0$ – Hubble constant, c – light speed, Z - redshift.

With that the absolute luminosity can be calculated in Sun luminosity units $L_{sun}$ through absolute magnitudes M by expressions

$$L/Lsun = 10^{(2/5)(Msun-M)}; \qquad (12)$$

$$M = m - 5 \lg (d/d_0); \qquad (13)$$

Where $M_{sun} \approx 4.7$, $d_0 = 10ps = 3.086 \ 10^{17}$ m.

As the luminosities of the galaxies lie in a range $10^7$-$10^{11}$Lsun, it is convenient to express them in billions of sun luminosities GLsun $\equiv 10^9$Lsun. For $H_0 = 2.28 \ 10^{-18}$ ($c^{-1}$) we obtain an expression to calculate the absolute luminosity of magnitude object m ≡ mag в величинах GLsun

$$L/GLsun = (75.86/10^9)10^{-(2/5)M}; \qquad (14)$$

where

$$M = mag - 5 \lg (4.258 \ 10^8 Z); \qquad (15)$$

For the curved space described by the ΛCDM model, the dependency obtains a more complex nature. In this case it can be calculated by the method [10] with the use of the data, brought there, about the dependency of the luminosity distance to the radiating object $D_L$ for the different values of the parameters of ΛCDM модели {$\Omega_M$, $\Omega_\Lambda$} on redshift. At that the luminosity L is related to the observed density of the radiation flow S (apparent luminosity l, S ≡ l) by the expression

$$D_L^2 \equiv L/4\pi S; \qquad (16)$$

where$D_L$ – luminosity distance to the object. In the case of flat Euclidean $D_L \equiv D_H = (c/H_0)$. With that the absolute luminosity of the object with magnitude m can be calculated through its absolute magnitude M by the formula

$$m = M + DM + K; \qquad (17)$$

whereK is k-correction, and DM is the distance modulus, defined by expression

$$DM = 5 \log (D_L/10pc); \qquad (18)$$

On the Picture 8 we show calculation dependencies L/GLsun(Z) for the curves Rmag_mean(Z), shown on the Picture 6, obtained by the expressions (14),(15) for the flat space.

On the Picture 9 we show identic curves obtained with the use of brought in [10] dependencies $(D_L/D_H)(Z)$ for {ΩM, $\Omega_\Lambda$}= (1;0) and {ΩM, $\Omega_\Lambda$}= (0.2;0.8) (ΛCDM model) without consideration of k-correction (K = 0).

On the Picture 10 we show the dependencies L/GLsun(Z), obtained by the expressions (14),(15) for the flat space, for the curves Rmag_mean(Z), shown on the Pictures 4,5.

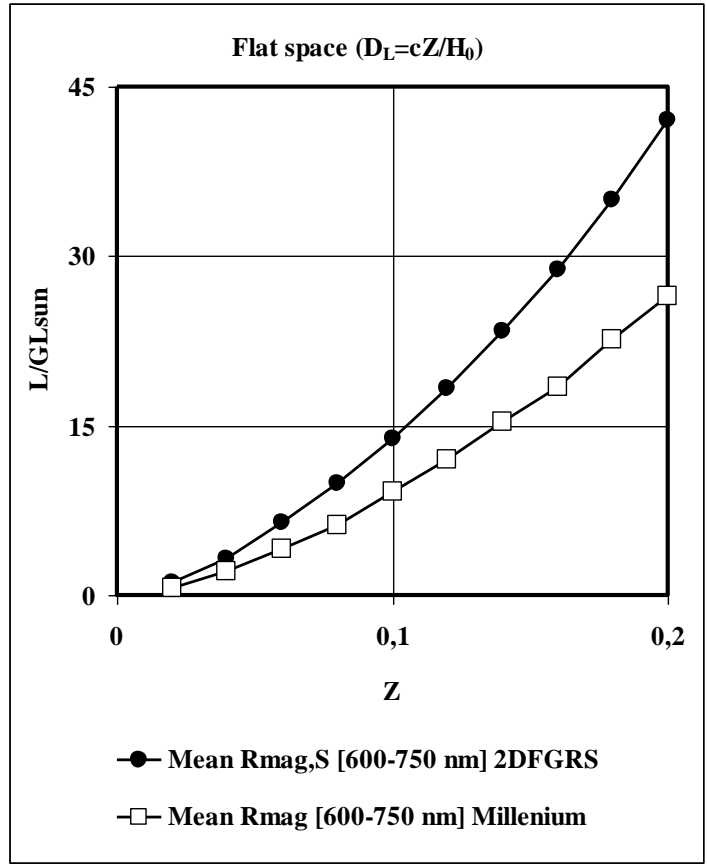

**Picture 8.** Calculation dependencies L/GLsun(Z) for the curves Rmag_mean(Z), shown on the Picture 6, for the flat space.

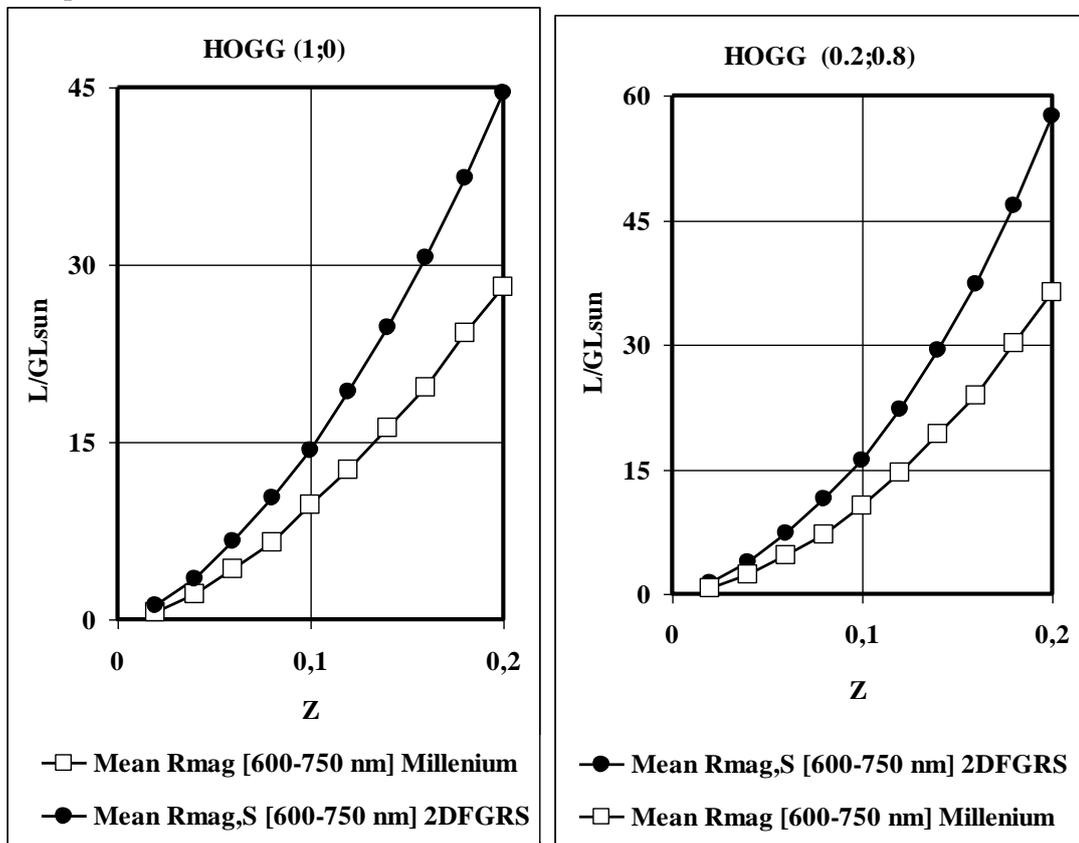

**Picture 9.** Calculation dependencies L/GLsun(Z) for the curves Rmag_mean(Z), shown on the Picture 6, obtained according to the data [10] for $\{\Omega M, \Omega_\Lambda\}$=(1;0) and $\{\Omega M, \Omega_\Lambda\}$=(0.2;0.8) ($\Lambda$CDM model).

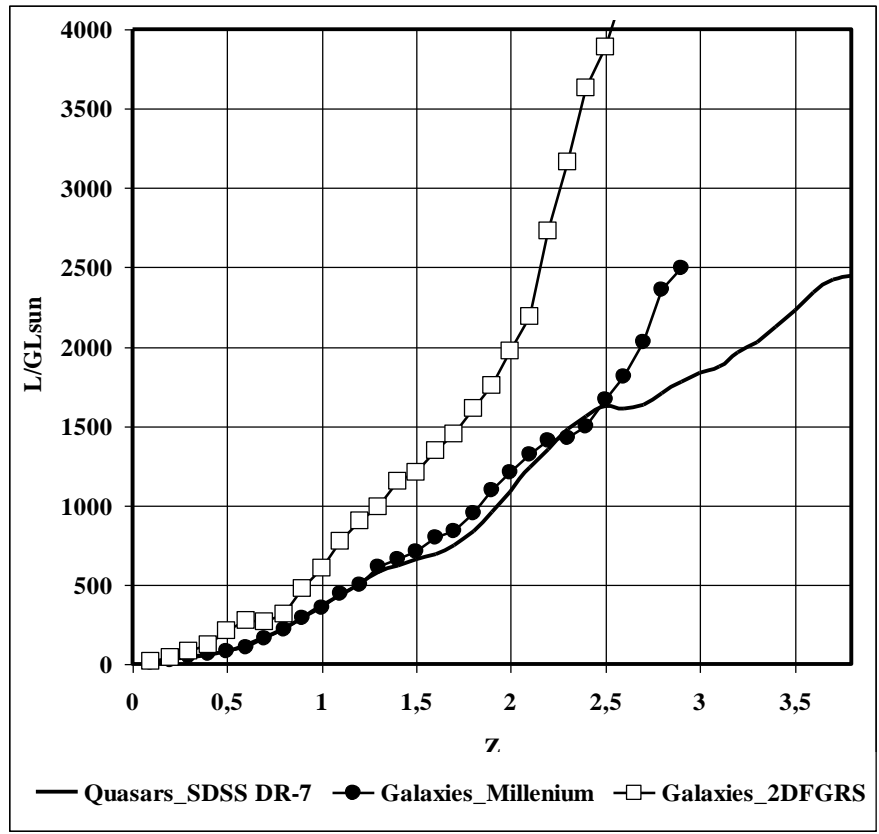

**Picture 10.** Dependencies on redshift of the absolute luminosity of the quasars and galaxies with magnitude values shown on the Pictures 4,5. Calculated by the expressions (14),(15) for the flat space.

The calculation results brought on the Pictures 8-10 show that the powers of the galaxies and quasars radiation, averaged by all the observed areas of the celestial sphere, located in the wide range of the observation angles (see Picture 11), according to the data of all the considered catalogs and all the calculation methods both for the flat space and ΛCDM model are increasing with the increase of the redshift.

In the Table 3 we bring the relations of the average radiation powers of the galaxies and quasars for the pairs of redshifts (0.02 and 0.2) and (0.2 and 2) according to the data on the Picture 10 and calculation results by the method [10].

| Catalog<br>Calculation method | Millenium (2003) | | 2DFGRS (2003) | | SDSSDR7 (2010) | |
|---|---|---|---|---|---|---|
| | $L_{0.02}/L_{0.2}$ | $L_{0.2}/L_{2.0}$ | $L_{0.02}/L_{0.2}$ | $L_{0.2}/L_{2.0}$ | $L_{0.02}/L_{0.2}$ | $L_{0.2}/L_{2.0}$ |
| **Flat space** | 47.4 | 46.7 | 38.3 | 48.0 | - | 37.1 |
| **HOGG (1;0)** | 48.3 | 68.2 | 39.0 | 70.1 | - | 53.7 |
| **ΛCDM<br>HOGG (0.2;0.8)** | 56.7 | 131.7 | 45.8 | 135.4 | - | 103.7 |

**Table 3.** Comparison of the average radiation powers of the distant observed objects (galaxies and quasars) for the pairs of redshifts (0.02 and 0.2) and (0.2 and 2.0), corresponding to the dependencies, shown on the Pictures 8,9,10.

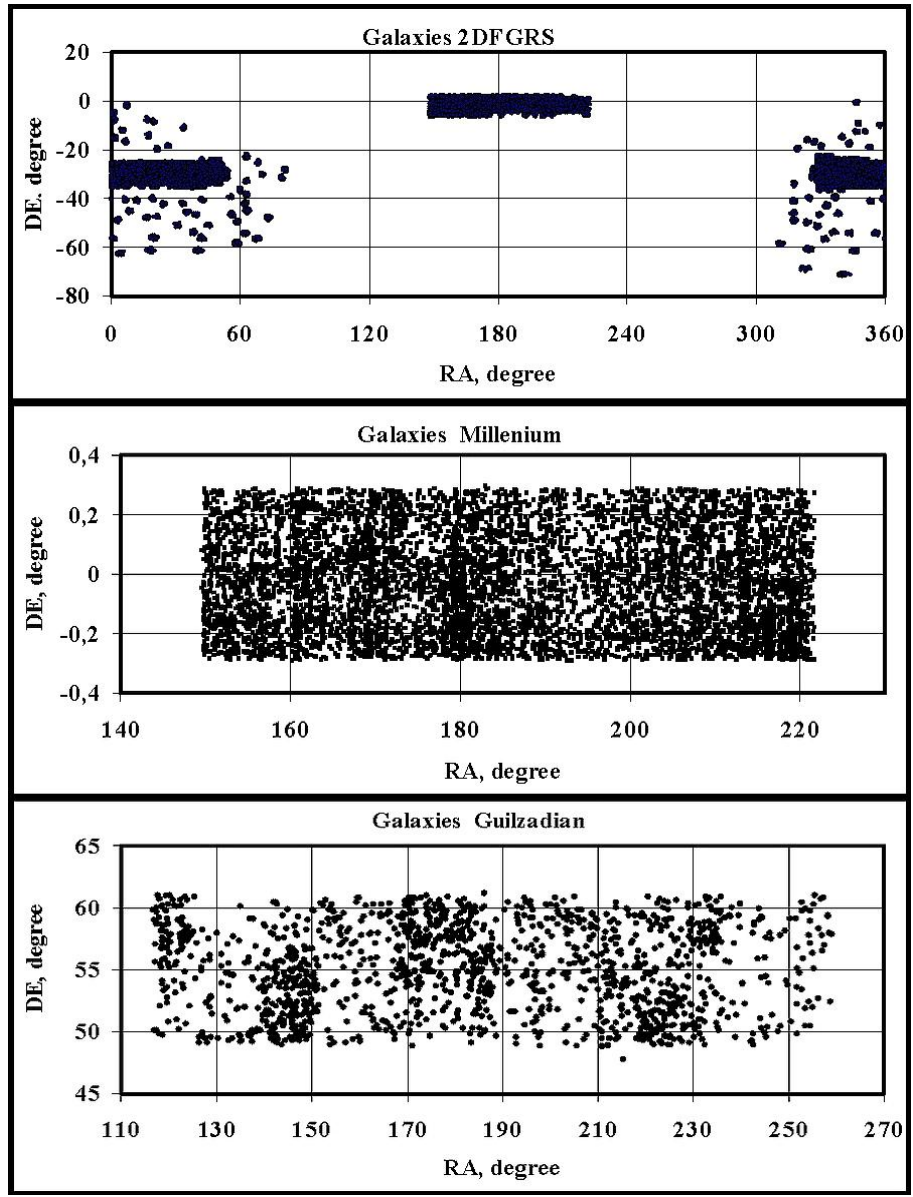

**Picture11.** Disposition of the galaxies on the celestial sphere [4],[5],[6].

When the redshift increases by 10 times the average radiation power of the object increases from 37 to 135 times, and the whole increase of the average power of the galaxies with increase of Z from 0.02 to 2.0 reaches 7 thousand times. Since the average data obtained for objects located at different parts of the sky, these changes can not be associated with any local area of space. One could assume that they are caused by the change of the average radiation power of objects in time. However, it seems more logical to assume that these results are related not to the real change of galaxies and quasars power in time but to the specifics of the used calculation methods. And because we obtained similar data using different calculation methods, possibly the matter is in the common model on which they are based.

It seems to be appropriate to analyze a possibility of the theoretical modeling of the obtained results within the modified cosmological model [11], built on basis of the metric tensor obtained with consideration of the non-zero value of the scale factor differential in the expanding Universe.

## 7. Calculation of the dependency of object's redshift on the angular coordinate in the modified cosmological model.

As we have shown in [11], in the metrics of the standard cosmological model there is an inner contradiction which can be eliminated if we build the metrics with consideration of the non-zero value of the scale factor differential in the expanding Universe. The modified model coming out from these metrics describes the Universe, being closed at any matter density, not requiring the introduction of any additional non-observed substances (cosmological constant or dark matter), infinite in time and at this development stage expanding in an accelerated manner.

The slower dynamics of its expansion eliminates the existing problems of the standard cosmological model related to the time restriction and also allows interpreting the recently discovered microwave background symmetry [12] as a natural phenomenon of the central symmetry of the celestial sphere not related to the isotropy violation and other fundamental principles of the relativity theory. The phenomenon of the central symmetry discovered in the effects of the microwave background central symmetry [13] and quasars [2], is related to the fact that in the modified model there is no characteristic for the standard model time restriction for the light beam movement which gives it a possibility to cover the distances exceeding the semi-circle of the closed Universe (l>πa). At that the light radiated by the source can reach the observer not only by the short straight way but also by the long arc of the big circle around the closed Universe, and be observed in the opposite point of the celestial sphere.

The 2-dimensional section of the 4-dimensional closed space with the hypersphere of radius «a» are shown on the Picture 12.

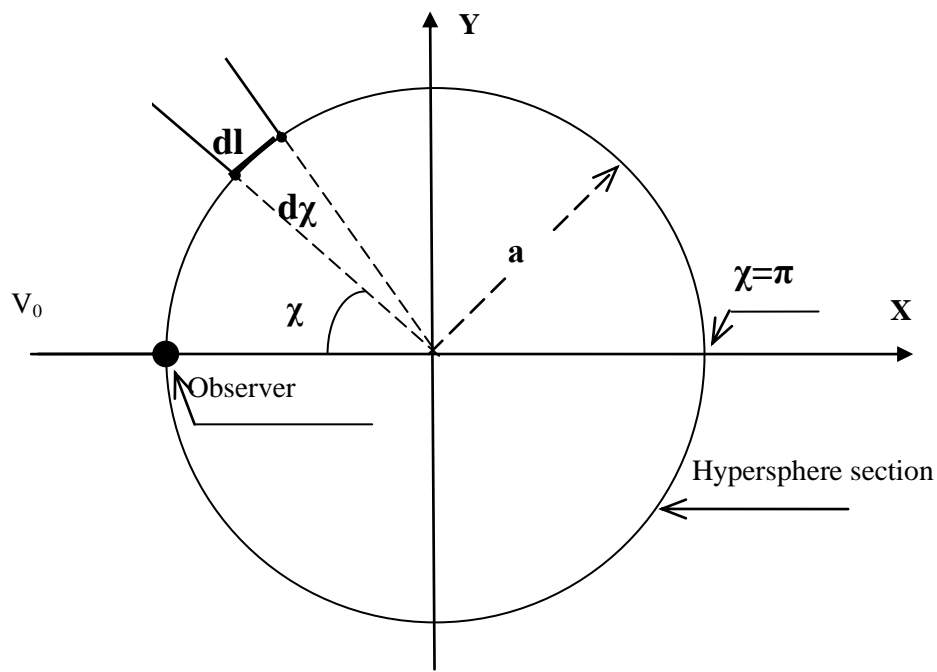

**Picture 12.** The 2-dimensional section of the 4-dimensional space with the hypersphere of the radius «a». The angular coordinate $\chi$ characterizes the moving away of the hypersphere point where the observed object is located, from the location of the observer (black spot). The distance from the object to the observer is $l=a\chi$.

Let us see the dependency of the observed objects' redshift on the angular coordinate $\chi$ in the modified model. The main parameters of the model [11] are defined by the expressions:

$$a'_{obs} = da / cdt_{obs} = \alpha (1-\alpha^2)^{1/2} ; \tag{36}$$

where, a – scale factor (hypersphere radius), $t_{obs}$ – time in the non-moving observer's coordinates, c – light speed, $\alpha$ – relative value of the scale factor $\alpha = a/2a_0$, $a_0 = const$. From where

$$dt_{obs} = (2a_0/c) \, d\alpha / [\alpha(1-\alpha^2)^{1/2}]; \tag{20}$$

$$\alpha(\tau_{obs}) = 1/ch(\tau_{obs}); \tag{21}$$

where

$$\tau_{obs} \equiv (c/2a_0)t_{obs}. \tag{22}$$

Reference to $\tau_{obs}$ is performed from the moment of the maximal expansion of the hypersphere ($\alpha = 1$).

Geometrical parameters of the Universe in the modified model are related to the relative density of the matter $\Omega$ and Hubble constant by the expressions:

$$\alpha(\Omega) = \Omega^{1/2}(\Omega+1)^{-1/2}; \tag{23}$$

$$a_0 = (c/2H)(1-\alpha^2)^{1/2} = (c/2H)(\Omega+1)^{-1/2}; \tag{24}$$

$$a = 2a_0\alpha = (c/H)\Omega^{1/2}(\Omega+1)^{-1}. \tag{25}$$

Increase of the distance from the object to the observer at small increase of the angular coordinate $d\chi$, equal to $ad\chi$, the light signal will cover for the time $dt_{obs}$

$$dl = ad\chi = -cdt_{obs}, \tag{26}$$

where from (20) we obtain

$$d\chi = - d\alpha / [\alpha^2(1-\alpha^2)^{1/2}]. \tag{27}$$

Integrating (27), we find the dependency $\chi(\alpha)$:

$$\chi(\alpha) = - \int d\alpha/[\alpha^2(1-\alpha^2)^{1/2}] = [(1-\alpha^2)^{1/2} / \alpha] + C_2. \tag{28}$$

Coefficient $C_2$ would be found from the condition: at $\chi = 0$ $\alpha = \alpha_0$. $C_2 = - [(1-\alpha_0^2)^{1/2}/\alpha_0]$, from where

$$\chi(\alpha) = [(1-\alpha^2)^{1/2}/\alpha] - [(1-\alpha_0^2)^{1/2}/\alpha_0]. \tag{29}$$

Considering that

$$Z = (\alpha_0/\alpha) - 1. \tag{30}$$

where $\alpha$ is related to the moment of the light radiation, $a\alpha_0$ – to the moment of its observation, i.e. to the present time, we obtain

$$\chi(Z) = [((Z+1)/\alpha_0)^2 - 1)]^{1/2} - [(1/\alpha_0)^2 - 1)]^{1/2}. \tag{31}$$

where the current value of $\alpha \equiv \alpha_0$ is related to the matter density $\Omega$ by the expression (23).

## 8. Dependency of the apparent luminosity on redshift in the modified cosmological model

The modified model describes the Universe as being closed at any matter density, i.e. being a 3-dimensional hypersphere expanding, as in the standard model, in a 4-dimensional Euclidean [8] space. The section area of the space on which the source radiation power is spread evenly, in the closed Universe is related to the angular coordinate $\chi$ of the source [14] by expression

S = 4πa²sin²χ;                                                            (32)

where a = const is the hypersphere radius at the moment of observation.

When the signal covers the distances corresponding to the angular coordinates of the sources χ>π/2, which in the modified model is possible because of the rather slow dynamics of the development, we can expect the appearance of the regularities, that are absent in the standard model, more precisely – the increase of the apparent luminosity of the source with the increase of the distance to it, when it corresponds to the decrease of the area S by the expression (32), which can appear in the average magnitude values dependency on redshift with negative slope areas corresponding to the increase of the apparent luminosity of the objects with the increase of the distance near Z ≈ 1 and Z ≈ 2.

Let us study the relation between the apparent luminosity of the object l and its absolute luminosity L (radiated power), which according to (32) can be described by a simple expression

l(Z) = L/S = L/4πa²sin²χ;                                          (33)

In an idealized variant of an absolutely homogenous Universe with zero velocities of the objects and strictly constant value of the space curvature radius all the beams from the source located inan angular distance χ = nπ, must converge in a spot, giving an infinite apparent luminosity. At that the dependency l(Z), expressed by the formula (33), will lead to the periodic dependency mag(Z), shown on the Picture 13, significantly different from the shown there dependencies mean_rmag(Z), obtained after the processing the observation data.

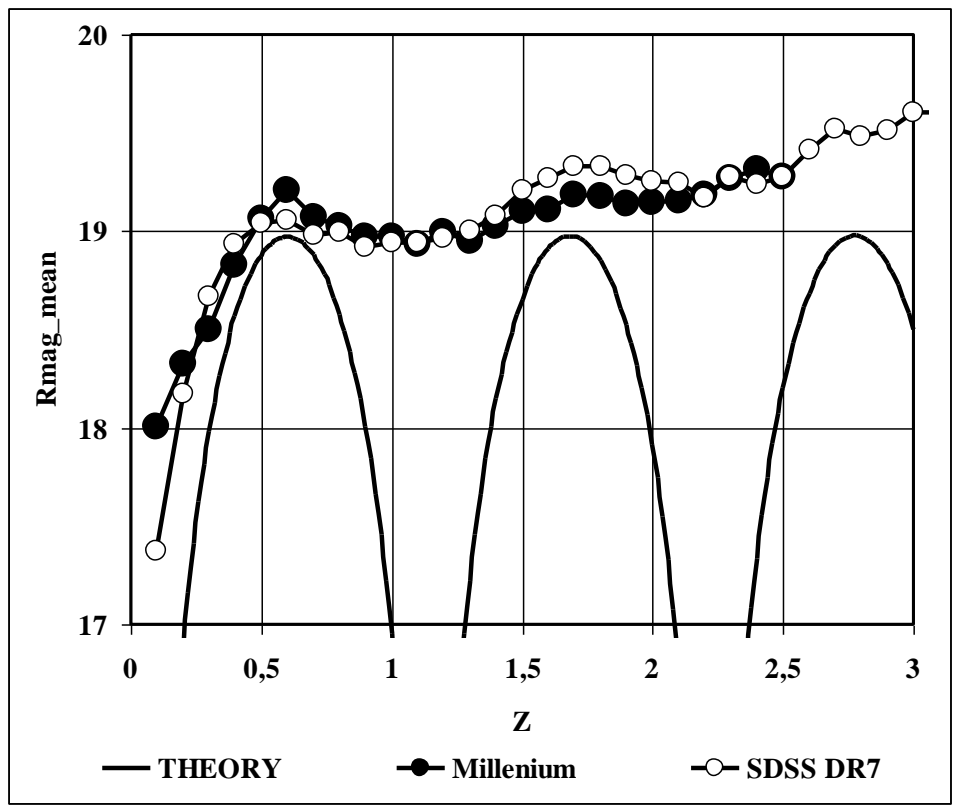

**Picture 13.** The comparison of the calculation dependency mag(Z) for l(Z) = L/4πa²sin²χ (33) at L=const with the dependencies Rmag_mean(Z) for the galaxies from the catalog Millenium (2003) and quasars from the catalog SDSS DR-7 (2010).

We can assume that the reason of the significant difference between the theoretical dependency and the real dependencies by the results of the observation data processing, shown on the Picture 13 is the non-compliance of the Universe to the ideal variant because of the dispersion and deviation of the radiations due to the non-constant space curvature, gravitational lensing, peculiar objects velocities, etc., which leads to the impossibility of the convergence of the beams radiated by a distant object in one spot of the space.

Insignificant linear growth of the Rmag_mean values with increase of Z can be related to the absorption of the electromagnetic waves in the intergalactic space. To consider it we can add an additional factor $\exp(-q\chi)$ to the expression (33), where q is the absorption coefficient:

$$l(Z) = L \exp(-q\chi)/4\pi a^2 \sin^2\chi. \tag{34}$$

## 9. Consideration of the influence of the non-ideal homogeneity of the space.

The non-correspondence of the Universe to the ideal variant with the strictly constant value of the space curvature can be described in the "fluffy" hypersphere model, where the beams, coming to the observer from the observed source go through the areas of the space with different values of the matter density corresponding to the different values of the curvature (scale factor).

With this purpose we have performed the numeric modeling of the addition of the radiation flows, that took place at the movement from the source to the observer through the areas of the space with different values of the matter density, hence with different values of the relative scale factor $\alpha$, changing according to the randomness law. By the expression (34) we have calculated the densities of the radiation flows $l_i(Z)$ for 99 independent channels (rays, spreading from the source located in an angular distance corresponding to the redshift Z).

On every step of beams movement ($\Delta Z=0.02$) we have calculated the growth of the angular coordinate $d\chi$ by the follow from (27),(30) expression

$$d\chi = dZ / [\alpha_0(1-\alpha^2)^{1/2}]; \tag{35}$$

for $\alpha(Z)$ of a homogeneous world in the words of (30).To account for the motion of rays in space with sharply varying matter density values $d\chi(Z)$ in each of the 99 channels were multiplied by a random value, calculated by the formula

$$d\chi_i = d\chi(Z)*K*random(0;1)^n; \tag{36}$$

for different values of coefficient K and exponent n, where «random(0;1)» is the random function MS Excel.

Picture 14 shows the simulation result of inhomogeneities by the expression (36). The nature of the plots $\chi_i(Z)$ in different channels is in qualitative agreement with the preferential movement of the beam in a space with a very low density of matter (the horizontal portions of the curves $\chi_i(Z)$) with randomly distributed small areas of high density (sharply growing parts of the curves).

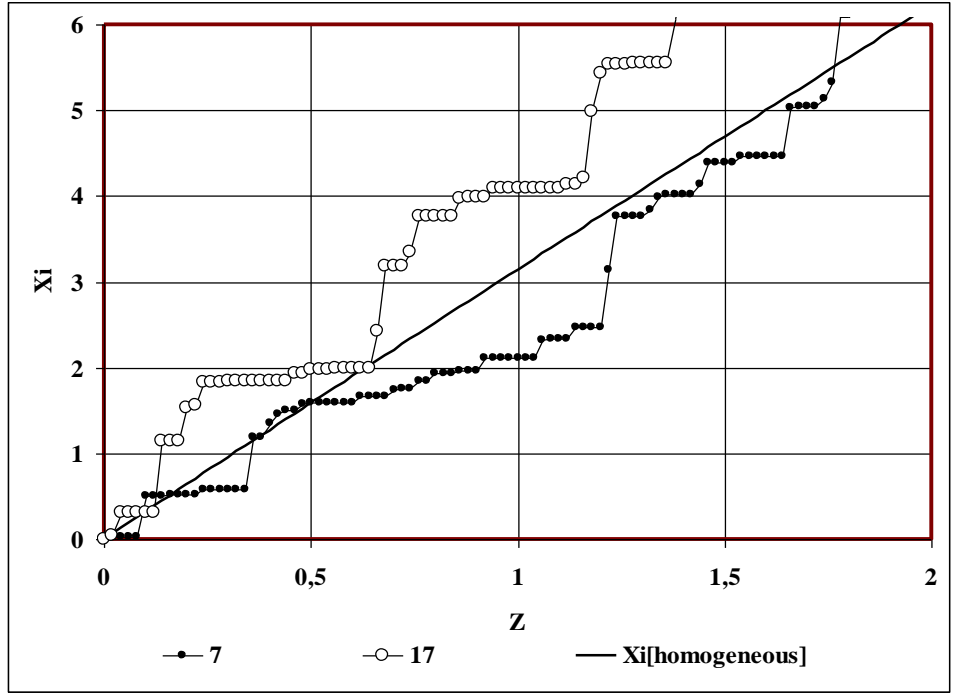

**Picture 14.** Modeling space heterogeneity by expression (18). The examples of dependency χi (Z) for 2 of the 99 channels (in this case channel #7 and #17) for n = 12, K = 15 in comparison with the dependence χ(Z) of a homogeneous universe.

Then for each of the 99 channels by summing the values $d\chi_i$ we calculated the dependencies $\chi_i(Z)$, corresponding to them radiation dispersion surface areas $S_i(Z) = 4\pi a^2 \sin^2\chi_i(Z)$, and individual densities of the radiation flows $l_i(Z)$. The individual densities of the flows $l_i(Z)$ were summed, giving the final value of the observed density of the radiation flow $l(Z) = \Sigma\ l_i(Z)$. The apparent magnitude mag(Z) was calculated from l(Z) by the common for the system (u,g,r,i,z) method «asinh» [7] with the parameters corresponding to the «r» range.

The obtained calculation curves appeared to be significantly depending on chosen parameters n and K, but having the form that is characterized by the same basic particularities as the dependencies obtained after the averaging of the observation data, i.e. having the converging oscillation nature with negative slope areas rmag(Z).

Picture 15 shows the dependencies of mag (Z), for different values of the exponent n. Coefficient K was chosen so as to ensure the average χi (Z), corresponding to approximately χ (Z) of a homogeneous universe (χi_mean (Z = 1) ≈ π). Calculated dependence for n = 18, compared with curves mean_rmag (Z) is shown in Picture 16.

Due to very high influence of the random component shows curves, averaging over 20 successive settlements with the new generation of random numbers. Shown curves are calculated for the values Ω = 0.12, H = 2.28 10⁻¹⁸ (с⁻¹), q = 0.08, for L = const(Z) ≈ 3*10¹⁰Lsun (for the convenience of joint demonstration of the curves slightly varied).

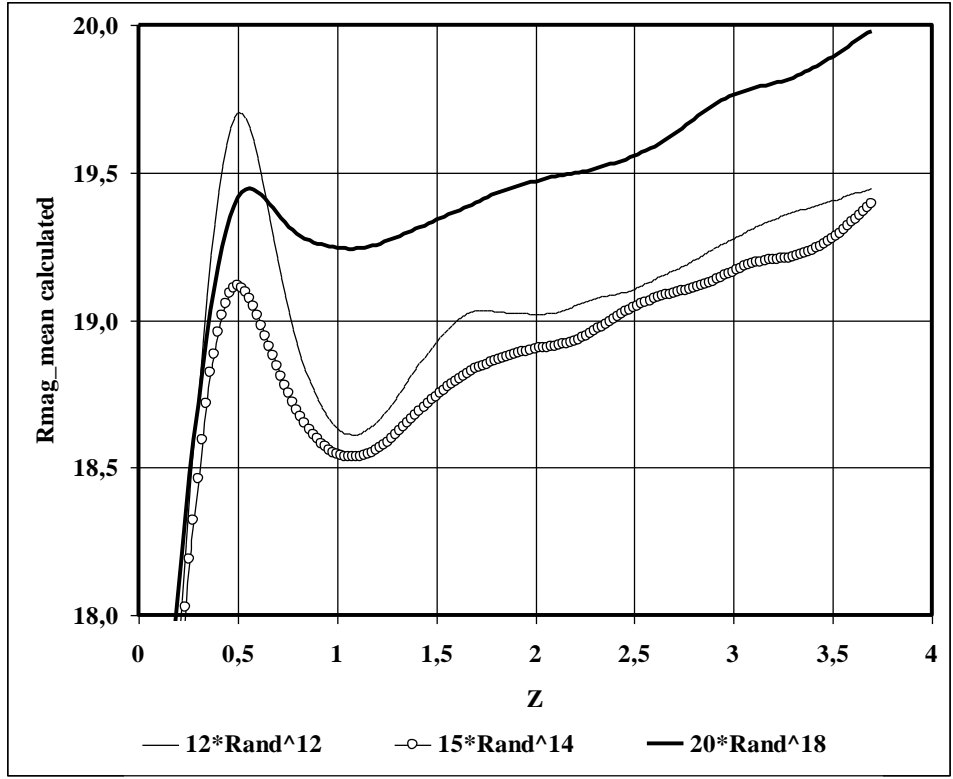

**Picture 15.** Calculated curves mag (Z) for different values of n and K. The oscillation amplitude decreases rapidly with increasing exponent n from 12 to 18.

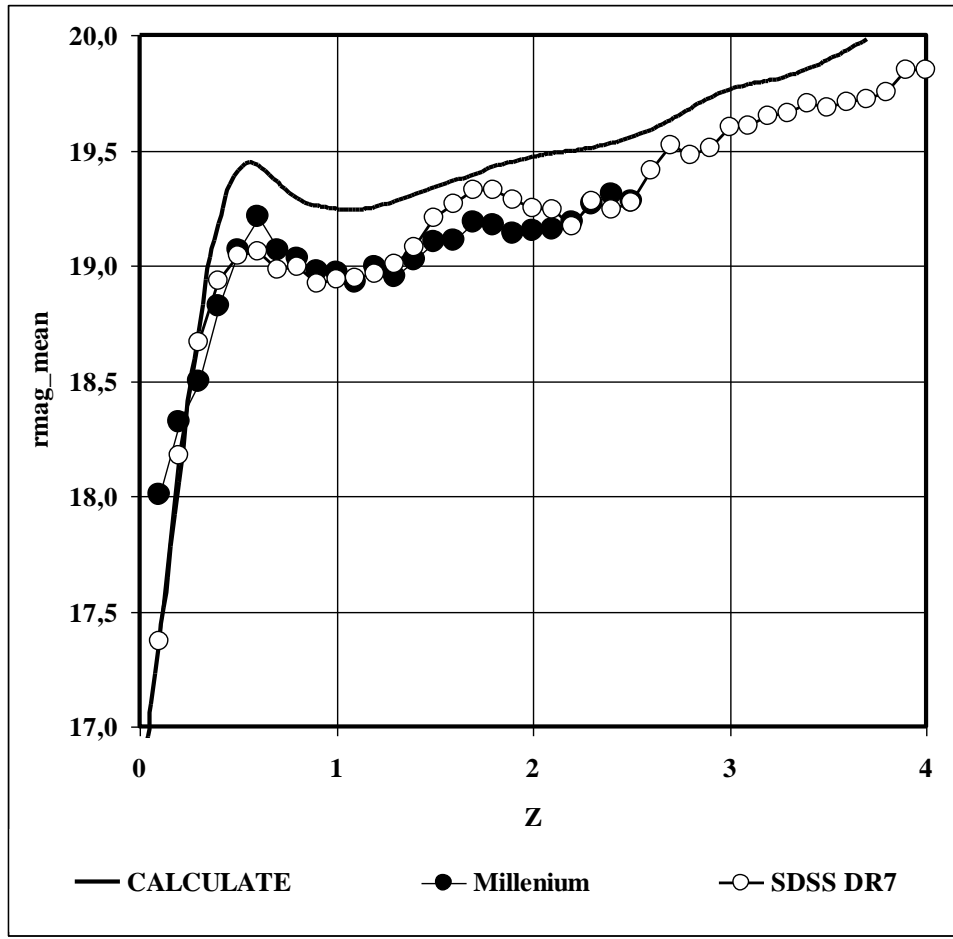

**Picture 16.** Comparison of the calculated curve mag (Z) with the dependencies Rmag_mean(Z) for the galaxies from the catalog Millenium (2003) and quasars from the catalog SDSS DR-7 (2010).

**10. Conclusion.**

The results of the comparative analysis of the data about the magnitudes of the quasars from the SDSS-DR5 catalog in different frequency ranges in the system (u,g,r,i,z) show that on the smoothed curves mag(Z) in frequency ranges u,g,r there are characteristic features both similar for all the curves and different for different curves. At Z<2 all the curves have practically the same form with two characteristic sections of the negative slope, at Z>2 the forms of the curves significantly differ. The nature of these differences can be interpreted as an influence of light absorption by the neutral hydrogen on Lyman-alpha line, manifesting on the curves in the ranges u,g and missing on the curve in the range r.

Then we have studied the dependencies on redshift of the averaged over the sky magnitude values in the range r, free from the Lyman-alpha absorption effect, for the galaxies from catalogs 2DFGRS, Millenium, Gyulzadian and quasars from catalog SDSS DR-7 (about 105 thousands of quasars and 175 thousands of galaxies). The results of the analysis show that for all the distant radiating objects, bith for quasars and galaxies, the dependencies of the magnitudes average values on Z are close by absolute values and form, and have a nature of converging oscillations with areas of negative slope near $Z \approx 1$ and $Z \approx 2$, corresponding to the increase of the apparent luminosity of the objects with the increase of the distance.

The results of the calculation of the average luminosities of the objects for the flat space and different variants of the $\Lambda$CDM model show that the radiation powers of galaxies and quasars, averaged by all the observed areas of the celestial sphere, are increasing severely at calculation both for the flat space and for the $\Lambda$CDM model with the increase of the redshift. The increase of the average power of the galaxies with the increase of Z from 0.02 to 2.0 reaches 7 thousand times.

It seems to be logical to assume that these results are related not to the real change of galaxies and quasars power in time but to the specifics of the used calculation methods. And because we obtained similar data using different calculation methods, possibly the matter is in the particularities of the metrics of standard cosmological model lying at their basis.

In the modified model built on the metrics which considers – in distinction from the standard model – the non-zero value of the scale factor differential in the expanding Universe, there may appear the regularities that are absent in the standard model, including the effect of the increase of the apparent luminosity of the sources with the increase of the distance. But in the idealized variant of the homogenous Universe with the strictly constant value of the space curvature radius, all the beams from the source located on an angular distance $\chi = n\pi$, must converge in a spot, giving the infinite apparent luminosity which does not happen in reality.

In relation to this, in this work we study the possibility of the numeric modeling of the dependency of distant radiating object magnitudes on redshift in the modified model, built on the beliefs about beam movement from the source to   the observer in the space with not constant but randomly changing curvature, defined at any step of beam movement by the pseudo-random number generator («random» function MS Excel).

We have shown that the calculation of the dependency of the distant radiating objects apparent magnitudes on redshift, performed by this methodology allows obtaining the theoretical mag(Z) for the object of constant luminosity L = const(Z), close to the average magnitudes of the quasars and galaxies dependencies on redshift, obtained  from the results of the observation data.

**Acknowledgments**

We are very grateful for the invaluable opportunity to use the SDSS information.

Funding for the SDSS and SDSS-II has been provided by  the Alfred P. Sloan Foundation,  the Participating Institutions, the National Science Foundation, the U.S. Department of Energy, the National Aeronautics and Space Administration, the Japanese Monbukagakusho,  the Max Planck Society, and the Higher Education Funding Council for England. The SDSS Web Site is http://www.sdss.org/.

The SDSS is managed by the Astrophysical Research Consortium for the Participating Institutions. The Participating Institutions are the American Museum of Natural History, Astrophysical Institute Potsdam, University of Basel, University of Cambridge, Case Western Reserve University, University of Chicago, Drexel University, Fermilab,  the  Institute for Advanced Study, the Japan Participation Group, Johns Hopkins University, the Joint Institute for Nuclear Astrophysics, the Kavli Institute for Particle Astrophysics and Cosmology, the Korean Scientist Group,  the Chinese Academy of Sciences  (LAMOST), Los Alamos National Laboratory, the Max-Planck-Institute  for  Astronomy  (MPIA),  the Max-Planck-Institute  for  Astrophysics (MPA),  New  Mexico State University, Ohio State University, University of Pittsburgh, University of Portsmouth, Princeton University, the United States Naval Observatory, and the University of Washington. SDSS Technical Paper References: [15-21].

*$25^{th}$ of December, 2012*